\documentclass[reqno]{amsart}
\usepackage{amsfonts}
\usepackage{amsthm}
\usepackage{amssymb}
\usepackage{dsfont}
\usepackage{amscd}
\usepackage{color}
\usepackage{tikz}
 \usetikzlibrary{calc,decorations.markings,matrix,decorations.pathmorphing}

\theoremstyle{definition}

\theoremstyle{remark}

\numberwithin{equation}{section}

\def\be{\begin{equation}}
\def\ee{\end{equation}}
\def\ba{\begin{eqnarray*}}
\def\ea{\end{eqnarray*}}
\def\bae{\begin{eqnarray}}
\def\eae{\end{eqnarray}}
\def\bc{\begin{center}}
\def\ec{\end{center}}

\begin{document}

\title[Hydrodynamical spectral evolution]{Hydrodynamical spectral evolution  for random matrices}
\author{Peter J. Forrester} \address{Department of Mathematics and Statistics, The University of Melbourne, Victoria 3010, Australia; ARC Centre of Excellence for Mathematical \& Statistical Frontiers}
\author{Jacek Grela} \address{M. Smoluchowski Institute of Physics and Mark Kac Complex Systems Research Centre, Jagiellonian University,  PL--30348 Krak\'ow, Poland}
%\date{}							

\begin{abstract}
The eigenvalues of the matrix structure $X + X^{(0)}$, where $X$ is a random Gaussian Hermitian matrix and
$X^{(0)}$ is  non-random or random independent of $X$, are closely related to Dyson Brownian motion.
Previous works have shown how an infinite hierarchy of equations satisfied by the dynamical correlations become triangular
in the infinite density limit, and give rise to the complex Burgers equation for the Green's function of the corresponding 
one-point density function. We show how this and analogous partial differential equations, for chiral, circular and Jacobi
versions of Dyson Brownian motion follow from a macroscopic hydrodynamical description involving the current density
and continuity equation. The method of characteristics gives a systematic approach to solving the PDEs, and in the
chiral case we show how this efficiently reclaims the characterisation of the global eigenvalue density for
non-central Wishart matrices due to Dozier and Silverstein. Collective variables provide another approach to deriving the
complex Burgers equation in the Gaussian case, and we show that this approach applies equally as well to chiral matrices.
We relate both the Gaussian and chiral cases to the asymptotics of matrix integrals.
\end{abstract}

\maketitle
%%%%%%%%%%%%%%%%%%%%%%%%%%%%%%%%%%%%%%%%%%%%%%%%%%%%%%%%%%
\section{Introduction}

One of the most basic questions in random matrix theory asks for the limiting global spectral density, given the distribution
on the elements of the matrices or the distribution on the space of matrices. Perhaps the most celebrated result in this class
is the Wigner semi-circle law. It applies to real random symmetric, or complex Hermitian matrices, in which the 
diagonal entries are independently distributed with mean zero and variance unity, and the 
upper triangular off-diagonal entries 
are independently distributed
with mean zero and variance two. 
Random matrices of this type are referred to as Wigner matrices. Consider a specific class of Wigner matrices
of size $N \times N$. A scaling of the matrices so that the eigenvalue density has compact support is
referred to as a global limit. For Wigner matrices this is achieved by
dividing each matrix by $\sqrt{2N}$
before taking the  $N \to \infty$ limit. Moreover, the corresponding
global spectral density $\rho^{\rm Wig}(x)$, normalised to integrate to unity, is given by
\begin{equation}\label{2A}
\rho^{\rm Wig}(x) = \left \{
\begin{array}{ll} \displaystyle {2 \over \pi} \sqrt{1 - x^2}, & |x| \le 1 \\[.2cm]
0, & |x| > 1.
\end{array} \right.
\end{equation}

In its fully generality, the most common proof of this result, and in fact the one provided by Wigner \cite{Wi55}, is to show that
\begin{equation}\label{2Ab}
\lim_{N \to \infty} \Big ( {\sqrt{2} \over N} \Big )^{2k} \langle {\rm Tr} \, X^{2k} \rangle = {1 \over k + 1} \Big ( {2k \atop k} \Big ),
\end{equation}
where the RHS is the $k$-th Catalan number. This is equivalent to establishing
 that the moments $\int_{-1}^1 x^{2k}
\rho^{\rm Wig}(x) \, dx $ are given by the $k$-th Catalan number for each $k=0,1,2,\dots$.
The functional form (\ref{2A}) now follows as a consequence of the Wigner semi-circle law being the unique
probability density with such moments.
 
An extension of the question seeking the global density of a single Wigner matrix
is to ask for
the global density of the sum of matrices
\begin{equation}\label{2X}
Y = X^{(0)} + X,
\end{equation}
where $X$ is again a Wigner random matrix, but $X^{(0)}$ may be non-random, or random independent of $X$. To answer this a tool kit beyond the analysis of moments is required. Specifically, the analytic properties of the Stieltjes transform plays an essential role.
Thus let $ X^{(0)}/  \sqrt{2N}$ have global spectral density $\rho^{(0)}(x)$, and define the corresponding  Green's function
 (or  Stieltjes transform) $G^{(0)}(z)$ by
\begin{equation}\label{G}
G^{(0)}(z) = \int_{I^{(0)}} { \rho^{(0)}(x) \over z - x} \, dx,
\end{equation}
where $I^{(0)}$ is the support of $ \rho^{(0)}$. Then, with $G(z)$ denoting the Green's function  of the sought global
density $\rho(y)$ of $Y/ \sqrt{2N}$, it is a known result (see e.g.~\cite[Th.~18.3.2]{PS11}) that
$G(z)$ is determined as the solution of the functional equation
\begin{equation}\label{4.1}
G(z) = G^{(0)}\Big (z - {1 \over 4} G(z) \Big ),
\end{equation}
subject to the requirements that $G(z) \sim  1/z$ as $|z| \to \infty$ and 
 that $G(z)$ be analytic for $z \notin I$, where $I$ is the support of $\rho(y)$.
 
As an illustration of (\ref{4.1}), suppose 
\begin{equation}\label{1.5a}
\rho^{(0)}(x) = \delta(x),
\end{equation}
the Dirac delta function at the origin,
in which case $G^{(0)}(z) =  1/z$. Then the functional equation
(\ref{4.1}) reads
$$
G(z) =  {1 \over z - G(z)/4},
$$
and this has the solution
\begin{equation}\label{5.1}
G(z) = 2( z - \sqrt{z^2 - 1}  ).
\end{equation}
But according to the Sokhotski-Plemelj formula for the inverse of the Stieltjes transform,
\begin{equation}\label{1.6a}
\rho(y) = {1 \over 2 \pi i} \lim_{\epsilon \to 0^+} \Big ( G(y - i \epsilon) -
G(y + i \epsilon)  \Big ), \qquad y \in I,
\end{equation}
and substituting (\ref{5.1}) reclaims (\ref{2A}).

Our interest is to develop
a viewpoint of the  functional equation (\ref{4.1}) for the Green's function of the density for the
matrix structure (\ref{2X}) as having origins in the
hydrodynamical equation
\begin{equation}\label{H}
{\partial \rho(x; \hat{\tau}) \over \partial \hat{\tau}} =   {\partial \over \partial x}  \bigg ( \rho(x;\hat{\tau})
{\partial \over \partial x} \Big ( V_1(x) + \int_{-\infty}^\infty  \rho(y;\hat{\tau}) V_2(x,y) \, dy \Big ) \bigg ),
\end{equation}
where $V_1(x)$, $V_2(x,y)$ are particular one and two body potentials, and $\hat{\tau}$ is a scaled parameter.
The scaling is required to compensate for the normalisation of $ \rho(y;{\tau})$ being such that its total integral
is unity rather than $N$. As to be reviewed in Section \ref{S2}, in the case that
$X$ in (\ref{2X}) is a standard Gaussian matrix, it is well known that a hydrodynamical equation,
in particular the complex Burger's equation, relates to (\ref{4.1}).

The advantage of viewing (\ref{4.1}) as a consequence of (\ref{H}) is that
a hydrodynamical description applies 
equally well to the case that
$X$ and $X^{(0)}$ have 
the chiral structure
\begin{equation}\label{H1}
X = \begin{bmatrix} 0_{m \times m} & Z_{m \times n} \\
Z_{n \times m}^\dagger & 0_{n \times n} \end{bmatrix}, \qquad n \ge m,
\end{equation}
and similarly $X^{(0)}$ with $Z$ replaced by $Z^{(0)}$, where 
$Z$ is a standard real or complex Gaussian matrix, and with $Z^{(0)}$ non-random or random independent of $Z$.
The class of random matrices (\ref{H1}) is fundamental to random matrix theory
(see e..g.~\cite[\S 3.1]{Fo10}). 
Although a hydrodynamical description of the spectral evolution of complex chiral Gaussian matrices has
been given in the recent works \cite{WARCHOLZAHED}, the general solution of the resulting partial differential equation
for the Green's function seems to have not been considered.
In Section \ref{S3} we show how (\ref{H}) leads to 
a partial differential equation, transformable to
an inhomogeneous complex Burger's equation. The partial differential equation was
first derived in \cite{G}, \cite{BN13},
with the starting point in the latter being a microscopic description involving Dyson Brownian motion
as reviewed in the introduction to Section \ref{S3}. Moreover, we show how to solve this partial differential
equation in a form analogous to (\ref{4.1}). In so doing we are able to reclaim a functional equation for
the Green's function of so-called non-central Wishart matrices given by Dozier and Silverstein \cite{DS}.

Specifically, consider random matrices of the form
$$
W = \sqrt{2 \hat{\tau} \over m} \tilde{Z} + Z^{(0)},
$$
where $\tilde{Z}$, $Z^{(0)}$ are rectangular $m \times n$ matrices, $\tilde{Z}$ a standard Gaussian
and $Z^{(0)}$ fixed or random independent of $\tilde{Z}$. Let $\hat{a} = \lim_{n,m \to \infty} (n/m) - 1$,
and let the $m,n  \to \infty$ limiting density of eigenvalues of $Z^{{(0)}\dagger} Z^{(0)}$ be equal to
$\rho^{\rm W, (0)}(y)$, and the limiting density of eigenvalues of $W^\dagger W$ be equal to
$\rho^{\rm W}(y;\hat{\tau})$. In this setting, and with
\begin{equation}\label{ggt}
g^{\rm W}(z;\hat{\tau}) = \int_{I_+} {\rho^{\rm W}(y;\hat{\tau}) \over z - y} \, dy,
\end{equation}
$I_+ \subset R^+$ denoting the support of $\rho^{\rm W}(y;\hat{\tau})$, we show that
\begin{equation}\label{ggW}
g^{\rm W}(z;\hat{\tau})  = F_W \int_{I_+^{(0)}} { \rho^{\rm W, (0)}(y) \over
z (F_{\rm W})^2 - 2 \hat{a} \hat{\tau} F_{\rm W} - y} \, dy,
\end{equation}
where $F_{\rm W} = 1 - 2 \hat{\tau} g^{\rm W}(z;\hat{\tau})$. With the identifications
$\hat{\tau} = \sigma^2 c/2$, $\hat{a} = (1 - c)/c$ and $g^{\rm W} = - m$ this is Eq.~(1.1) in \cite{DS}. 

When both $X$ and $X^{(0)}$ in \eqref{2X} are unitary, we are dealing with circular ensembles for which a hydrodynamical equation \eqref{H} is well known \cite{PS91}.
We review the necessary working in Section \ref{S4} and proceed to present a novel hydrodynamical description of Jacobi ensemble in trigonometric variables, which can be interpreted as a "chiral" deformation of circular ensembles. In Section \ref{S5} and \ref{S6}, we turn to the collective variables approach which is used to review the well-known asymptotic expansion of Harish-Chandra/ Itzykson--Zuber integral and to study not previously  considered expansion of 
Berezin--Karpelevich type integrals.

%%%%%%%%%%%%%%%%%%%%%%%%%%%%%%%%%%%%%%%%%%%%%%%%%%%%%%%%%%
\section{Gaussian ensembles}\label{S2}
In this section we review how the hydrodynamical equation (\ref{H}) comes about from the study of the
eigenvalues of the matrix sum (\ref{2X}) in the case that $X$ is a standard Gaussian matrix with real ($\beta = 1$)
or complex ($\beta = 2$) entries \cite{Dy72}, and furthermore leads to the functional equation
(\ref{4.1}). The starting point is the fact that the Gaussian distribution on the space of
matrices $\{Y\}$ which are real symmetric $(\beta = 1)$ or complex Hermitian $(\beta = 2)$,
\begin{equation}\label{P1}
P_\tau(X^{(0)};Y) = {1 \over C_{N,\tau}} \exp \Big ( - {\beta \over 4 \tau} \text{Tr} (Y - X^{(0)})^2 \Big ),
\end{equation}
where $C_{N,\tau}$ denotes the normalisation constant, satisfies the diffusion equation
\begin{equation}\label{mu}
{\partial P_\tau \over \partial \tau} = {1 \over \beta} \sum_\mu D_\mu {\partial^2 P_\tau \over \partial Y_\mu^2}.
\end{equation}
In (\ref{mu}) the label $\mu$ ranges over the independent elements, including both the real and imaginary parts of the
off-diagonal elements if they are complex, and $D_\mu = 1$ for the diagonal elements, and $D_\mu = {1 \over 2}$
for the off-diagonal elements.

There is also a class of Hermitian matrices with distribution (\ref{P1}) that satisfy (\ref{mu}) with $\beta = 4$.
Thus the Hermitian matrix $Y$, and similarly $X^{(0)}$, is now a $2N \times 2N$ matrix formed from an
$N \times N$ matrix with each element a $2 \times 2$ matrix of the form
$$
\begin{bmatrix} z & - w \\ - \bar{w} & z \end{bmatrix}, \qquad z,w \in \mathbb C.
$$
Such $2 \times 2$ matrices are isomorphic to the real quaternion division algebra, one of only three
associative real normed division
algebras along with the real and complex numbers, so in this case $Y$ is said to have real quaternion entries;
see e.g.~\cite[\S 1.3.2]{Fo10}.

The differential operator on the RHS of (\ref{mu}) can be interpreted as the Laplace-Beltrami operator $\nabla^2$
associated with the
metric form for the matrix spaces
$$
(ds)^2 = {\rm Tr} \, (d Y dY^\dagger) = \sum_{\mu,\nu} g_{\mu \, \nu} dY_\mu dY_\nu,
$$
with $\mu$ labelling the independent elements as in (\ref{mu}) and similarly
$\nu$, and where $g_{\mu \, \nu}  = {1 \over D_\mu} \delta_{\mu,\nu}$; see e.g.~\cite[eq.~(11.9)]{Fo10}.
Introducing the diagonalisation formula for $Y$, $Y = U L U^\dagger$, where $U$ is the matrix of eigenvectors and $L$ is
the diagonal matrix of the eigenvalues, allows $\nabla^2$ to be rewritten according to the separated form
\begin{equation}\label{N1}
\nabla^2 = {1 \over J} \sum_{j=1}^N {\partial \over \partial \lambda_j}
\Big ( J {\partial \over \partial \lambda_j} \Big ) + O_U, \qquad J = \prod_{j < k } |\lambda_k - \lambda_j|^\beta,
\end{equation}
where the operator $O_U$ involves derivatives with respect to variables relating to the eigenvectors only.
The significance of this is that the eigenvalue distribution $p_\tau(\lambda_1,\dots,\lambda_N)$ obtained by integrating over the angles $U$ and the
distribution $P_0(X^{(0)})$ of $X^{(0)}$ in (\ref{P1}),
\begin{equation}\label{P2}
p_\tau(\lambda_1,\dots,\lambda_N) = J \int dU \int dX^{(0)} P_\tau \left (X^{(0)};U L U^\dagger \right ) P_0\left (X^{(0)} \right ) 
\end{equation}
satisfies the Smoluchowski-Fokker-Planck equation
\begin{equation}\label{FP}
{\partial p_\tau \over \partial \tau} = {\mathcal L} p_\tau, \qquad
{\mathcal L}  = \sum_{j=1}^N {\partial \over \partial \lambda_j} \Big ( {\partial W \over \partial \lambda_j} +
\beta^{-1}{ \partial \over \partial \lambda_j} \Big ),
\end{equation}
with
\begin{equation}\label{W}
W = - \sum_{1 \le j < k \le N} \log | \lambda_j - \lambda_k|.
\end{equation}
In the case that the $\tau$ dependence in (\ref{P1}) is modified so that $P_\tau$ satisfies not (\ref{mu}), but the
heat equation for Brownian motion in a harmonic potential i.e.~the Smoluchowski-Fokker-Planck equation (\ref{FP}) with $W$
correspondingly modified by the addition of an harmonic potential ${1 \over 2} \sum_{j=1}^N \lambda_j^2$,
was first derived by Dyson \cite{Dy62}. As such the corresponding process is referred to as Dyson Brownian motion.

As pointed out in \cite{Dy62}, the Smoluchowski-Fokker-Planck equation (\ref{FP}) has a more standard interpretation than its origin in
random matrix theory. Specifically, consider a classical system of $N$ particles interacting on a line with potential $W$.
Suppose the particles execute overdamped Brownian motion in a fictitious background fluid with friction coefficient $\gamma$,
and furthermore the system is at inverse temperature $\beta$. It is a basic fact --- see e.g.~\cite{Ri92} --- that in this setting
the evolution of the probability density $p_\tau(\lambda_1,\dots,\lambda_N)$ for the location of the particles at
positions $\lambda_1,\dots\lambda_N$ is given by the Smoluchowski-Fokker-Planck equation (\ref{FP}), where the LHS is to be multiplied by
the friction coefficient $\gamma$. The random matrix problem gives rise to the specific potential (\ref{W}), corresponding
to the particles interacting via the repulsive pair potential $V_2(x,y) = - \log|x - y|$, and so the underlying classical gas is
referred to as a log-gas \cite{Fo10}.

Our interest is in the one-body dynamical density $\rho(x;\tau)$ defined as an average over \eqref{P1},
\begin{align}
\label{rhodef}
	\rho(x;\tau) = \frac{1}{N} \left < \sum_{i=1}^N \delta(x - \lambda_i) \right >_{P_\tau},
\end{align}
which has been normalized to have total integral unity. We probe a global regime for which the eigenvalues are scaled so that they have
finite support as in (\ref{2A}). Note that since the integral over $x$ of
$ \rho(x;\tau)$ is unity, the integral of $N \rho(x;\tau)$ over $x$ is $N$.
In this scaling the inter-particle spacing goes to zero, and the response of the system to perturbation
is governed by macroscopic equations \cite{Dy72}. Relevant to the one-body dynamical density is the macroscopic equation
\begin{equation}
N^2J(x;\tau) = \mathcal{F}(x;\tau),
\end{equation}
where $N^2J(x;\tau)$ is the one-body current related to the density by the continuity equation
\begin{equation}\label{C}
{\partial \over \partial \hat{\tau}} \rho(x;\hat{\tau}) = - {\partial \over \partial x} J(x;\hat{\tau}), \qquad \hat{\tau} = N \tau
\end{equation}
(the scaled parameter $\hat{\tau}$ is introduced to compensate for the integral of
$ \rho(x;\tau) $ normalised to unity)
while $ \mathcal{F}(x;\tau)$ refers to the macroscopic force density. For the log-gas in the long wavelength regime
the force density to leading order is of an electrostatics origin, so implying (see e.g.~\cite{FJ96})
\begin{equation}
J(x;\tau) = - \rho(x;\tau)  {\partial \over \partial x} \Big ( - \int_{-\infty}^\infty \rho(x';\tau)
\log | x - x'|  \, dx' \Big ). 
\end{equation}
Differentiating both sides with respect to $x$, and making use of the continuity equation (\ref{C}) on the RHS, we see
the hydrodynamical equation (\ref{H}) results with $V_1(x)=0$ and $V_2(x,y) = - \log | x - y|$.

We now want to show how this particular hydrodynamical equation leads to the functional equation (\ref{4.1}). 
For this purpose, it is convenient to introduce the Hilbert (or Cauchy) transform as the principal value integral
\begin{equation}\label{Ht}
\mathcal{H}[v](x) :={\rm PV} \int_I {v(y) \over x - y} \, dy, \qquad x \in I.
\end{equation}
The hydrodynamical equation of interest then reads
\begin{equation}\label{H12}
{\partial \rho(x; \hat{\tau}) \over \partial \hat{\tau}} =   - {\partial \over \partial x}  \bigg ( \rho(x;\hat{\tau})
\mathcal{H}[\rho(\cdot;\hat{\tau})](x)  \bigg ).
\end{equation}
We also introduce the Green's function
\begin{equation}\label{GG}
G(z;\hat{\tau}) := \int_I {\rho(y;\hat{\tau}) \over z - y} \, dy, 
\end{equation}
cf.~(\ref{G}).

Next we follow the working in \cite[\S III.B.4]{Be96}, which begins by noting that as a consequence 
of the residue theorem, the
Green's function is related to the
Hilbert transform by
\begin{equation}\label{gm}
G_\pm(x;\hat{\tau}) = \mp i \pi \rho(x;\hat{\tau}) + \mathcal{H}[\rho(\cdot;\hat{\tau})](x), \qquad x \in I.
\end{equation}
Using this in (\ref{H12}) gives
\begin{equation}
\label{gpm}
2 {\partial \over \partial \hat{\tau}} (G_-(x;\hat{\tau}) - G_+(x;\hat{\tau})) =
- {\partial \over \partial x} ((G_-(x;\hat{\tau}))^2 - (G_+(x;\hat{\tau}))^2) .
\end{equation}
It must therefore be that the function
$$
2 {\partial \over \partial \hat{\tau}} G(z;\hat{\tau})  +
{\partial \over \partial z} (G(z;\hat{\tau}))^2 
$$
is analytic throughout the entire complex plane. But according to (\ref{GG}), $G(z;\hat{\tau}) \sim 1/z$ as 
$|z| \to \infty$, so this function furthermore goes to zero at infinity. The only analytic function with this property
is the zero function, and so after minor manipulation we have
\begin{equation}\label{G2}
 {\partial \over \partial \hat{\tau}} G(z;\hat{\tau})  +
 G(z;\hat{\tau})  {\partial \over \partial z} G(z;\hat{\tau})
= 0.
\end{equation}
This is the Euler equation, or equivalently complex Burgers equation of hydrodynamics. Thus with
$z = x + i y$ and $G(z;\tau) = U + i V$, $(U,V)$ is the velocity field at point $(x,y)$ in the plane for an ideal
fluid at constant pressure. 

To solve the initial value problem for \eqref{G2}, we invoke the method of complex characteristics where both $z$ and $G$ are complex functions. This is a slight generalization of a standard technique applicable to inital value problems of real first order PDEs. We present this method for a general first order equation of the form
\begin{align}
\label{geng}
	A\Big (G(z;\hat{\tau}),z,\hat{\tau} \Big ) \frac{\partial}{\partial \hat{\tau}} G(z;\hat{\tau}) + 
	B\Big (G(z;\hat{\tau}),z,\hat{\tau} \Big ) \frac{\partial}{\partial z} G(z;\hat{\tau}) = 
	C\Big (G(z;\hat{\tau}),z,\hat{\tau}
	\Big ).
\end{align}
The main idea is to seek  a coordinate transform $(z,\hat{\tau}) \to (\alpha,\beta)$ such that the PDE \eqref{geng} becomes an ODE along the curves of constant $\alpha$,
\begin{align}
	\label{cond}
	\frac{d}{d\beta} G(z;\hat{\tau}) = C\Big (G(z;\hat{\tau}),z,\hat{\tau} \Big )
\end{align}
called characteristic lines or simply characteristics.
By the chain rule $\frac{d}{d\beta} = \frac{d \hat{\tau}}{d \beta} {\partial \over \partial \hat{\tau}} + \frac{d z }{d \beta} {\partial \over \partial z}$, the left-hand sides of \eqref{cond} and \eqref{geng} dictate the system of equations describing the characteristics,
   \begin{align}
   \label{chars}
   	& \frac{d}{d \beta}\hat{\tau}(\alpha,\beta) = 
	A\Big (G(\alpha,\beta),z(\alpha,\beta),\hat{\tau}(\alpha,\beta)\Big ), \nonumber \\
	& \frac{d}{d \beta}z(\alpha,\beta) = 
	B\Big (G(\alpha,\beta),z(\alpha,\beta),\hat{\tau}(\alpha,\beta) \Big ),
\end{align}
where $G(\alpha,\beta) = G(z(\alpha,\beta);\hat{\tau}(\alpha,\beta))$. These equations form lines in the $(z,\hat{\tau})$ space, labeled by the $\beta$ parameter and passing through the prescribed initial point $(z(\alpha,0), \hat{\tau}(\alpha,0))$. For the latter to be determined, the Green's function $G(z;\hat{\tau}=0)$ on the $\hat{\tau}=0$ line is required. With this initial data specified, the set of equations \eqref{cond} and \eqref{chars} are in principle solvable by standard means and comprise the sought solution to \eqref{geng}. 

The equation \eqref{G2} is an instance of \eqref{geng} with $A=1,B=G$ and $C=0$.
We read off from (\ref{geng}) and (\ref{chars}) that the 
differential equations describing characteristic lines and the propagation of the solution are
\begin{align}
	\frac{d}{d \beta}z(\alpha,\beta) & = G(\alpha,\beta), \qquad \frac{d}{d \beta}\hat{\tau}(\alpha,\beta) = 1, \qquad \frac{d}{d \beta}G(\alpha,\beta) = 0.
\end{align}
The initial data comprises of the initial position $z(\alpha,0) = \alpha, \hat{\tau}(\alpha,0) = 0$ and the starting Green's function $G(\alpha,0)=G(z(\alpha,0);\hat{\tau}(\alpha,0)) = G(\alpha;0)$. Explicit integration gives
\begin{align}
	G(\alpha,\beta) & = G(\alpha;0), \nonumber \\
	\hat{\tau}(\alpha,\beta) & = \beta, \nonumber \\
	z(\alpha,\beta) & = \alpha + \beta G(\alpha;0).
\end{align}
These, after eliminating $\alpha$ and $\beta$, yield the functional equation
\begin{equation}\label{G3}
 G(z;\hat{\tau}) = G \Big ( z - \hat{\tau}  G(z;\hat{\tau}) ;0 \Big ),
 \end{equation}
 which is to be compared to (\ref{4.1}). Equivalently, recalling the definition
\eqref{G}, we obtain the  implicit integral equation
\begin{align}
	G(z;\hat{\tau}) = \int_{I^{(0)}} \frac{\rho^{(0)}(\mu)d\mu}{z- \hat{\tau}G(z;\hat{\tau}) - \mu}.
\end{align}
Working closely related to the above discussion can be found in \cite{Wa14}.

To anticipate the precise relationship between
 (\ref{G3}) and (\ref{4.1}), let us choose
 $\tau =1/(4N)$. The RHS of \eqref{P1} is then proportional to $\exp(-N\beta(Y - X^{(0)})^2)$, and thus we see that
 $Y = X + X^{(0)}$, where $X =  \tilde{X} / \sqrt{2N}$ with $\tilde{X}$ a standard Gaussian matrix. This is precisely
 the setting which gives rise to (\ref{4.1}). On the other hand, the choice  $\tau =1/(4N)$ is,
 according to (\ref{C}), equivalent to the choice $\hat{\tau} = 1/4$, and this substituted in (\ref{G3}) gives
 (\ref{4.1}).
 
  In the Introduction the functional equation (\ref{4.1}) was illustrated by
 showing that the case $\rho^{(0)}(x) = \delta(x)$ leads to the Wigner semi-circle law (\ref{2A}).
 Another example which permits an explicit functional form for the density is when
 \begin{equation}\label{Ga2}
 G^{(0)}(z) = {1 \over 2} \Big ( {1 \over z - a} + {1 \over z + a} \Big ).
 \end{equation}
 This corresponds to an initial density
 \begin{equation}\label{Ga3}
 \rho^{(0)}(x) = {1 \over 2} ( \delta(x-a) + \delta(x+a) ),
  \end{equation}
 or equivalently to $X^{(0)}$ in (\ref{2X}) being a diagonal matrix with half its eigenvalues at $a$
 and the other half at $-a$. Substituting (\ref{Ga2}) in (\ref{G3}) with $\hat{\tau} = 1$
 shows that the Green's
  function $G(z)$ satisfies the cubic equation
 \begin{equation}\label{Ga4}
 G^3 - 2z G^2 + (1 - a^2 + z^2)G - z = 0.
 \end{equation}
 With $\xi = z - G$, this equation first appeared  in the present context in \cite{BK}, where it was shown
 to correspond to a spectral density supported on two disjoint intervals symmetrical about
 the origin for $a>1$. In the case $a=1$ the intervals meet at the origin, and it can then be shown that
 the eigenvalue density has the explicit form \cite[eq.~(6.118)]{Na11}
 \begin{equation}\label{G3b}
 \rho(y) =  \frac{ |y|^{1/3}}{2\sqrt{3}\pi}\Big((3\sqrt{3}+ \sqrt{27-8y^2})^{2/3}-(3\sqrt{3}- \sqrt{27-8y^2})^{2/3}\Big)    
 \end{equation}
for $-\sqrt{27/8}\leq y \leq \sqrt{27/8}$. For a discussion of (\ref{Ga4}) in terms of caustics corresponding to
the complex Burger's equation, see \cite{Wa14}.

 %%%%%%%%%%%%%%%%%%%%%%%%%%%%%%%%%%%%%%%%%%%%%%%%%%%%%%%%%%
 \section{Chiral Gaussian ensembles}\label{S3}
 \subsection{Partial differential equation}
 We now turn our attention to the matrix sum (\ref{2X}) in the case that $X$ is a chiral Gaussian random matrix
 as specified by (\ref{H1}), and $X^{(0)}$ has the same block structure as $X$ but is non-random or random
 independent of $X$. Thus the matrix $Y$ in (\ref{2X}) similarly has the block structure
 \begin{equation}\label{H1a}
Y = \begin{bmatrix} 0_{m \times m} & W_{m \times n} \\
W_{n \times m}^\dagger & 0_{n \times n} \end{bmatrix}, \quad W = Z + Z_0.
\end{equation}
Instead of (\ref{P1}), we now have a distribution on the block matrix $W$ specified by
 \begin{equation}\label{P1a}
P_\tau(Z^{(0)};W) = {1 \over C_{N,\tau}^{\rm c}} \exp \Big ( - {\beta \over 4 \tau} \text{Tr} (W - Z^{(0)})^\dagger  (W - Z^{(0)}) \Big ).
\end{equation}

The diffusion equation (\ref{P1}), but with $Y$ replaced by $W$, again allows for a characterisation of this matrix
distribution. Before the (Hermitian) matrix $Y$ was decomposed according to the diagonalisation formula. The
appropriate decomposition of the matrix $W$ is the singular value decomposition $W = U L V^\dagger$, where $U$ and $V$
are real orthogonal ($\beta = 1$) or complex unitary ($\beta = 2$) matrices of size $m \times m$ and $n \times n$
respectively, while $L = {\rm diag} (x_1,\dots,x_m)$, where $\{x_j\}$ are the singular values of $W$ or equivalently
$\{x_j^2\}$ are the eigenvalues of $W^\dagger W$.
As discussed in \cite{Fo97,AW96}, integrating over the distribution of $Z^{(0)}$ leaves a distribution function $p_\tau(x_1^2,\dots,x_m^2)$ depending
on the parameter $\tau$ and the eigenvalues $W^\dagger W$ only. To characterise this distribution as the solution
of an evolution equation, we require the fact (see e.g.~\cite[\S 11.2.2]{Fo10}) that the Jacobian $J$ in the formula
(\ref{N1}) should now read
$$
J = \prod_{j=1}^m x_j^{\beta a + 1} \prod_{1 \le j < k \le m} | x_k^2 - x_j^2|^\beta, \qquad a = n - m + 1 - 2/\beta,
$$
and furthermore  on the RHS the replacements
\begin{equation}\label{R}
\{\lambda_j \} \mapsto \{x_j\}, \qquad N \mapsto m
\end{equation}
should be made.
Doing this then gives  that $p_\tau(x_1^2,\dots,x_m^2)$ satisfies the
Smoluchowski-Fokker-Planck equation (\ref{FP}) with 
\begin{equation}\label{W8}
W = - {a' \over 2} \sum_{j=1}^m \log x_j^2 - \sum_{1 \le j < k \le m} \log | x_k^2 - x_j^2|,
\end{equation}
where $a' = a + 1/\beta$ and with the replacements (\ref{R}). In the log-gas analogy, the domain is
now the half line $x > 0$, and there is both a one and two body potential given by
\begin{equation}\label{W9}
V_1(x) = - {a' \over 2}  \log x^2, \qquad V_2(x,y) = - \log |x^2 - y^2|.
\end{equation}

We now turn our attention to the hydrodynamical description of the global density $\rho^{\rm c}(x;\hat{\tau})$ which is defined as
\begin{align}
\label{rhochdef}
	\rho^{\rm c}(x;\hat{\tau}) = \frac{1}{m} \left < \sum_{i=1}^m \left ( \delta (x-x_i) + \delta (x+x_i) \right ) \right >_{{P_\tau}} ,
\end{align}
where we average over the measure \eqref{P1a}, and the superscript ``c" denotes the chiral case.
 This density is an even function in $x$ and is normalized so that integration over the positive half line $x>0$ gives unity. As in the discussion of Section \ref{S2}, to access the global regime
we must scale the parameter $\hat{\tau} = m \tau$, and also scale
\begin{equation}\label{As}
\hat{a} = \lim_{\substack{n\to \infty \\ m \to \infty}} {n \over m} -1,
\end{equation}
which so determines the limiting ratio $n/m$. In terms of these scaled parameters,
from the explicit form (\ref{W9}) of the one and two body potentials, and the fact that
the domain is a half line, the hydrodynamical equation (\ref{H}) reads
\begin{align}\label{Ha}
{\partial \rho^{\rm c}(x; \hat{\tau}) \over \partial \hat{\tau}} &=   {\partial \over \partial x}  \bigg ( \rho^{\rm c}(x;\hat{\tau})
{\partial \over \partial x} \Big (  - {\hat{a} \over 2}  \log x^2 -  \int_0^\infty  \rho^{\rm c}(y;\hat{\tau}) \log |x^2 - y^2| \, dy \Big ) \bigg )
\nonumber \\
& = 
 {\partial \over \partial x}  \bigg ( \rho^{\rm c}(x;\hat{\tau})
{\partial \over \partial x} \Big (  - {\hat{a} \over 2}  \log x^2 -  \int_{-\infty}^\infty  \rho^{\rm c}(y;\hat{\tau}) \log |x - y| \, dy \Big ) \bigg ),
\end{align}
where the second line follows by writing $\log|x^2 - y^2| = \log |x - y| + \log |x+y|$ and the fact that $\rho^{\rm c}(y;\hat{\tau})$  is
even in $y$.

Introducing the Hilbert transform as defined in (\ref{Ht}), (\ref{Ha}) can be written
\begin{equation}\label{Hb}
{\partial \rho^{\rm c}(x; \hat{\tau}) \over \partial \hat{\tau}}  =
- {\partial \over \partial x}  \bigg ( \rho^{\rm c}(x;\hat{\tau})
\mathcal{H}[\hat{a} \delta(\cdot) + \rho^{\rm c}(\cdot;\hat{\tau})](x)  \bigg ).
\end{equation}
Introducing too the Green's function
\begin{equation}\label{Hc}
G^{\rm c}(z;\hat{\tau}) = \int_{I_+ \cup - I_+} {\rho^{\rm c}(y;\hat{\tau}) \over z - y} \, dy,
\end{equation}
where $I_+ \subset \mathbb R^+$ is the support of $\rho^{\rm c}(y;\hat{\tau})$ on the positive real axis,
we see that
\begin{equation}\label{Hd}
G^{\rm c}_{\pm}(x;\hat{\tau}) = \mp i \pi \rho^{\rm c}(x;\hat{\tau}) +
\mathcal{H}[ \rho^{\rm c}(\cdot;\hat{\tau})](x), \qquad x \in I_+ \cup - I_+,
\end{equation}
where $G^{\rm c}_{\pm}$ is defined according to (\ref{gm}).
Proceeding as in the derivation of (\ref{G2}), it follows from the use of (\ref{Hd}) in (\ref{Hb}) that
\begin{equation}\label{He}
 {\partial \over \partial \hat{\tau}} G^{\rm c}(z;\hat{\tau}) - \frac{\hat{a}}{z^2} G^{\rm c}(z;\hat{\tau}) +
 \Big ( G^{\rm c}(z;\hat{\tau})  + {\hat{a} \over z} \Big ) {\partial \over \partial z} G^{\rm c}(z;\hat{\tau})
= 0 .
\end{equation}

\subsection{Solution of the partial differential equation}
We now seek the general solution of the initial value problem for  this partial differential equation, first in the case
$\hat{a} = 0$, then in the more difficult case $\hat{a} > 0$.
%%%%%%%%%%%%%%%%%%%%%%%%%%%%%%%%%%%%%%%%%%%%%%%%%%%%%%%%%%
\subsection*{The case $\hat{a} = 0$}
In the case $\hat{a} = 0$ we see that (\ref{He}) reduces to the Euler equation (\ref{G2}). Thus as with (\ref{G3}),
the solution in this case is
\begin{equation}\label{G3a}
 G^{\rm c}(z;\hat{\tau}) = G^{\rm c} \Big ( z - \hat{\tau}  G^{\rm c}(z;\hat{\tau}) ;0 \Big ),
 \end{equation}
 but we must keep in mind that $G^{\rm c}$ is defined by (\ref{Hc}) rather than (\ref{GG}). 
 For (\ref{G3}) and (\ref{G3a}) to imply identical, up to a scale factor,
 eigenvalue distributions
 and singular value distributions respectively for general $\hat{\tau}$, we see that we must have the initial conditions related by 
\begin{equation}\label{G3bb}
\rho(y;0) =  {1 \over 2} \rho^{\rm c}(y;0), \quad y \in I_+^{(0)} \cup - I_+^{(0)},
\end{equation}
where $I_+^{(0)}$ is the support of $\rho^{\rm c}(y;0)$ on the positive real axis.
The factor of ${1 \over 2}$ is to compensate for the
normalisation of the LHS being such that integration over the whole real line gives unity, while on the RHS integration
of $\rho^{\rm c}(y;0)$ over the half line $y  > 0$ gives unity which is evident from the definition \eqref{rhochdef}.

Specifically, we see that with the initial conditions related by (\ref{G3bb}), the solutions of (\ref{G3}) and
(\ref{G3a}) are related by
$$
G(z;\hat{\tau}) = {1 \over 2} G^{\rm c}\left (z;{\hat{\tau} \over 2} \right )
$$
and thus
\begin{equation}\label{W1}
\rho(y;\hat{\tau}) = {1 \over 2} \rho^{\rm c} \left (y;{\hat{\tau} \over 2} \right ).
\end{equation}
In relation to the initial condition $\rho(y;0)  = \delta(y)$, after recalling that the Wigner semi-circle law
(\ref{2A}) results from the parameter value $\hat{\tau} = 1/4$, we see from (\ref{W1}) that
\begin{equation}\label{G3c}
\rho^{\rm c}(x;\hat{\tau}) =
{1 \over 2 \pi \hat{\tau}} \sqrt{8 \hat{\tau} - x^2}, \qquad 0 \le  x \le 2 \sqrt{2\hat{\tau}}.
\end{equation}
It must therefore be that for $Z^{(0)} = 0_{m \times n}$ in (\ref{P1a}) the global density of the singular values of
the $m \times n$, $ n > m$, matrix $W$ with distribution specified by (\ref{P1a}), is in the case that 
$\lim_{m \to \infty} {n \over m} = 1$ equal to this functional form. 

To see this latter result, which is well known,
first note that
each singular value $x$ of $W$ is related to an eigenvalue $y$ of $W^\dagger W$ by $x^2 = y$.
Changing variables according to this prescription in  (\ref{G3c}) gives
the density function
\begin{equation}\label{G3d}
{1 \over 4 \pi \hat{\tau} \sqrt{y}} \sqrt{8\hat{ \tau} - y}, \qquad 0 < y  < 8 \hat{ \tau} .
\end{equation}
With $\hat{\tau}= {1 \over 8}$ this specifies the Marchenko-Pastur law for the limiting density of the
eigenvalues of the scaled matrices ${1 \over 4 m} W^\dagger W$, with $W$ now a standard Gaussian
rectangular matrix, again in the circumstance that
$\lim_{m \to \infty} {n \over m} = 1$; see e.g.~\cite[Section 3.4.1]{Fo10}.
This scaling is consistent with that implied by (\ref{H1a}), with
$Z^{(0)} = 0_{m \times n}$ and $\tau = \hat{\tau}/m = 1/(8m)$.

An analogous discussion holds for the initial density (\ref{Ga2}). In the case $a=1$, $\hat{\tau}=1/2$, and upon
changing variables $y^2 = x$, we conclude that the limiting density of the
eigenvalues of the scaled matrices ${1 \over m} (X +Z^{(0)})^\dagger (X+Z^{(0)})$, where 
$X$ is a standard Gaussian $m \times n$
rectangular matrix
$Z^{(0)}$ is
an $m \times n$ matrix with half its entries on the diagonal equal to +1,
the other half $-1$, all other entries equal to 0, is equal to
\begin{equation}\label{X1}
\frac{1}{2^{5/3}3^{1/2}\pi}\frac{(3\sqrt{3}+ \sqrt{27-4x})^{2/3}-(3\sqrt{3}- \sqrt{27-4x})^{2/3}}{x^{1/3}},
\end{equation}
supported on $0<x \leq \frac{27}{4}$. 
This is the density function for the Raney distribution with parameters $p=3$, $r=2$ \cite{pz}, where for
general $0 < r \le p$ and $p>1$ the Raney distribution is characterised by its $k$-th moments
according to
 \be  \label{raneynumber}
 R_{p,r}(k)=\frac{r}{pk+r}\binom{pk+r}{k}, \quad k=0,1,2,\dots .
 \ee
 Note that $R_{2,1}$ corresponds to the Catalan numbers; recall (\ref{2Ab}).
%%%%%%%%%%%%%%%%%%%%%%%%%%%%%%%%%%%%%%%%%%%%%%%%%%%%%%%%%%
\subsection*{The case $\hat{a} > 0$}
For nonzero parameter $\hat{a}$, the solution of \eqref{He} to an initial value problem is given by the method of characteristics described in Section \ref{S2}. The chiral ensemble is an instance of \eqref{geng} with $A=1,B=G^{\rm c}+\hat{a}/z$ and $C=\hat{a}G^{\rm c}/z^2$. Accordingly, the system of ODEs describing both the characteristic lines and the propagation of Green's function is
\begin{align*}
	& \frac{d}{d \beta}z(\alpha,\beta) = G^{\rm c}(\alpha,\beta) + \frac{\hat{a}}{z(\alpha,\beta)}, \\
	& \frac{d}{d \beta}G^{\rm c}(\alpha,\beta) = \hat{a} \frac{G^{\rm c}(\alpha,\beta)}{z(\alpha,\beta)^2}, \\
	& \frac{d}{d \beta}\hat{\tau}(\alpha,\beta) = 1,
\end{align*}
with initial conditions $z(\alpha,0) = \alpha, \hat{\tau}(\alpha,0) = 0$ and $G^{\rm c}(\alpha,0) = G^{\rm c}(z(\alpha,0);\hat{\tau}(\alpha,0)) = G^{\rm c}(\alpha;0)$. The last equation for $\hat{\tau}$ is simply solved as $\hat{\tau} = \beta$ whereas the first two are coupled but readily solved to give
\begin{align}
	z(\alpha,\beta) & = \sqrt{\alpha + G^{\rm c}_0(\alpha)\beta} \sqrt{\alpha+G^{\rm c}_0(\alpha) \beta + 2 \hat{a} \frac{\beta}{\alpha}} , \qquad \hat{\tau}(\alpha,\beta) = \beta, \\
	G^{\rm c}(\alpha,\beta) & = G^{\rm c}_0(\alpha) \frac{\sqrt{\alpha+G^{\rm c}_0(\alpha) \beta + 2 \hat{a} \frac{\beta}{\alpha}}}{\sqrt{\alpha + G^{\rm c}_0(\alpha)\beta}} ,
\end{align}  
where we used a simplified notation $G^{\rm c}_0(\alpha) = G^{\rm c}(\alpha,0)$.
 Now we make the substitution $\hat{\tau} = \beta$ and calculate auxillary formulas by multiplying the equations for $z$ and $G^{\rm c}$ and squaring the equation for $z$,
\begin{align}
\frac{G^{\rm c}(z;\hat{\tau})}{z} & = \frac{G^{\rm c}_0(\alpha)}{\alpha + G^{\rm c}_0(\alpha) \hat{\tau}}, \label{1st} \\
z^2 \alpha & = (\alpha + G^{\rm c}_0(\alpha) \hat{\tau})^2 \alpha + 2 \hat{a} \hat{\tau} (\alpha + G^{\rm c}_0(\alpha) \hat{\tau}). \label{2nd}
\end{align}
From \eqref{1st} we find $G^{\rm c}_0 = \frac{\alpha G^{\rm c}}{z-G^{\rm c}\hat{\tau}}$ and from that 
\begin{align}
\label{agt}
	\alpha + G^{\rm c}_0 \hat{\tau} = \frac{\alpha z}{z - G^{\rm c}\hat{\tau}},
\end{align}
where from now on we suppress the arguments of $G^{\rm c}$ and $G^{\rm c}_0$ for brevity.
We plug \eqref{agt} into \eqref{2nd} to obtain a formula for $\alpha^2$,
\begin{align}
\label{alphasq}
	\alpha^2 = (z-G^{\rm c}\hat{\tau})^2 - \frac{2\hat{a}\hat{\tau}(z-G^{\rm c}\hat{\tau})}{z}.
\end{align} 
Next we recall the definition  \eqref{Hc}
to determine $G^{\rm c}_0$ and use the symmetry of the initial spectral density $\rho^{\rm c, (0)}$ to rewrite
\begin{align}
\label{chiralf}
	G^{\rm c}_0(\alpha) = \int_{I_+^{(0)} \cup -I_+^{(0)}} \frac{\rho^{\rm c, (0)}(\mu)d\mu}{\alpha-\mu} = 2\alpha \int_{I_+^{(0)}} \frac{\rho^{\rm c, (0)}(\mu)d\mu}{\alpha^2-\mu^2}.
\end{align} 
This variant of $G^{\rm c}_0$  plugged into \eqref{1st}  gives
\begin{align}
	G^{\rm c} = \frac{z-G^{\rm c}\hat{\tau}}{\alpha } G^{\rm c}_0(\alpha) = 2(z-G^{\rm c}\hat{\tau}) \int_{I_+^{(0)}} \frac{\rho^{\rm c, (0)}(\mu)d\mu}{\alpha^2-\mu^2},
\end{align}
where we also used  \eqref{agt}. Lastly, we recall the equation \eqref{alphasq} for $\alpha^2$ and so
obtain an implicit integral equation
\begin{align}
\label{chfin}
 G^{\rm c}(z;\hat{\tau}) = 2 \int_{I_+^{(0)}} \frac{\rho^{\rm c,(0)}(\mu) d \mu}{z - G^{\rm c}(z;\hat{\tau})\hat{\tau} - \frac {2 \hat{a}\hat{\tau}}{z} - \frac{\mu^2}{z-G^{\rm c}(z;\hat{\tau})\hat{\tau}}} .
\end{align}
We can check, upon recalling (\ref{Hc}), that in the case $\hat{a} = 0$ (\ref{chfin}) is equivalent to (\ref{G3a}).

As a first illustration of (\ref{chfin}), let $\rho^{\rm c, (0)}(\mu) = 2 \delta(\mu)$, Then (\ref{chfin}) simplifies to
a quadratic for $G^{\rm c}$, and from this use of the 
Sokhotski-Plemelj formula (\ref{1.6a}) implies the density of singular values is equal to
\begin{equation}\label{3.16a}
{1 \over \pi x} \Big ( - x^4 + (2 \hat{a} + 4)x^2 - \hat{a}^2 \Big )^{1/2},
\end{equation}
where we have chosen $\hat{\tau} = 1/2$
as in deriving (\ref{X1}), supported on the region of the positive real axis such that the
argument of the square root is positive. This result is well known; see e.g.~\cite[Prop.~3.4.1]{Fo10}.

As a second illustration,
 suppose $\rho^{\rm c, (0)}(\mu) = \delta(\mu-b) + \delta(\mu+b)$ so that 
 the singular values of $Z^{(0)}$ are all located at $b$.
 We then find that  \eqref{chfin} gives a cubic equation for $G^{\rm c}$,
 \begin{equation}\label{3.30}
 g ( z^2 (1 - g)^2 - \hat{a} (1 - g) - b^2) = 1 - g,
 \end{equation}
%\begin{align}
%	G^{\rm c} \left ((z-G^{\rm c}\hat{\tau})^2 - \frac{2 \hat{a} \hat{\tau}}{z}(z-G^{\rm c}\hat{\tau}) - b^2 \right ) = 2(z-G^{\rm c}\hat{\tau}).
%\end{align}
where we have set $\hat{\tau} = 1/2$, and $g:=G^{\rm c}\hat{\tau}/z$. For $\hat{a} > 0$, we can see from
this that the density is supported away from the origin. The reasoning is that otherwise, for small
$z$, $g$ must behave like $z^{-1-\alpha}$ with $0 < \alpha < 1$, as would follow from its
relation to $G^{\rm c}$ and (\ref{Hc}).  But this is incompatible with
(\ref{3.30}) unless $\hat{a} = 0$.

%%%%%%%%%%%%%%%%%%%%%%%%%%%%%%%%%%%%%%%%%%%%%%%%%%%%%%%%%%
\subsection*{Equivalent viewpoints}
The Green's function (\ref{Hc}) is the Stieltjes transform of the density of the singular values of the matrix
$W$ in (\ref{H1a}). The singular values of $W$ also appear as the eigevanlues of the block matrix
$Y$ specified in (\ref{Hc}). Thus $Y$ has $n-m$ zero eigenvalues, $m$ eigenvalues equal to the singular
values of $W$, and $m$ eigenvalues equal to minus the singular values of $W$; see e.g.~\cite[Prop.~3.1.1]{Fo10}.
Denoting the corresponding density, normalised to integrate to unitary, by $\rho^{\rm ch}(x;\tau)$ we
see that
\begin{equation}\label{rr}
 \rho^{\rm ch}(x;\tau) = {1 \over 2 + \hat{a}} \rho^{\rm c}(x;\tau) + {\hat{a} \over 2 + \hat{a}}
 \delta(x).
 \end{equation}
 Hence, the corresponding Green's function
 $$
 g^{\rm ch}(z;\tau) := \int_{I_+ \cup - I_+} {\rho^{\rm ch}(x;\tau) \over z - x} \, dx
 $$
 is related to the Green's function (\ref{Hc}) by
 \begin{align}
\label{chwish}
g^{\rm ch}(z;\hat{\tau}) =  {1 \over 2 + \hat{a}}
G^{\rm c}(z;\tau) + {\hat{a} \over 2 + \hat{a}} {1 \over z}.
\end{align}
Substituting in 
 \eqref{He} gives the inhomogeneous Burger's equation
\begin{align}
\label{chiral}
	 \frac{\partial}{\partial \hat{\tau}} g^{\rm ch}(z;\hat{\tau}) + (2+\hat{a}) g^{\rm ch}(z;\hat{\tau}) \frac{\partial}{\partial z} g^{\rm ch}(z;\hat{\tau}) + \frac{\hat{a}^2}{(2+\hat{a})}\frac{1}{z^3} = 0,
\end{align}
which is the form considered in \cite{WARCHOLCHIRAL} for the present setting.

Another variant is to consider the eigenvalues of $W^\dagger W$, with $W$ as in (\ref{H1a}).
As already remarked below (\ref{P1a}), the eigenvalues are $W^\dagger W$ are the squared singular
values of $W$. Thus, with the corresponding density denoted by $\rho^{\rm W}(X;\tau)$, we have
\begin{equation}
\rho^{\rm W}(X;\tau) = {1 \over 2 \sqrt{X}} \rho^{\rm c}(\sqrt{X};\tau).
\end{equation}
The Green's function for the eigenvalues of $W$ is given in terms of $\rho^{\rm W}(X;\tau)$ by
$$
g^{\rm W}(z;\hat{\tau}) = \int_{I_+} {\rho^{\rm W}(x;\hat{\tau}) \over z - x} \, dx.
$$
Recalling (\ref{Hc}) we thus have the relation
\begin{align}\label{kG}
	g^{\rm W}(z^2;\hat{\tau}) = \frac{G^{\rm c}(z;\hat{\tau})}{2z}.
\end{align}
Substituting in  \eqref{He} gives that $g^{\rm W}$ satisfies the Burger's like equation
\begin{align}
\label{wish}
	\frac{\hat{a} + 1}{2} \frac{\partial}{\partial {\hat{\tau}}} g^{\rm W}(z;\hat{\tau}) + \left ( \hat{a} + 2 z 
	g^{\rm W}(z;\hat{\tau}) \right )\frac{\partial}{\partial z} g^{\rm W}(z;\hat{\tau}) + g^{\rm W}(z;\hat{\tau})^2 = 0.
\end{align}
This variant, modulo some rescaling, is the one given in \cite{G,BN13} in the present context.

For completeness we present the solution to the initial value problem of both \eqref{chiral} and \eqref{wish}, which
follow from  (\ref{chfin}) by substitution. In
the case of  $g^{\rm ch}$, substituting (\ref{rr}) and (\ref{chwish}) in
(\ref{chfin}) we find
\begin{align}
	g^{\rm ch}(z;\hat{\tau}) = 2 F_{\rm ch} \int_{I_+^{(0)}} \frac{\rho^{\rm ch,(0)}(\mu) d \mu }{F_{\rm ch}^2 - \frac{F_{\rm ch}\hat{a} \hat{\tau}}{z} - \mu^2} + \frac{\hat{a}}{2 + \hat{a}} \left ( \frac{1}{z} - \frac{2z}{zF_{\rm ch} - \hat{a}\hat{\tau}} \right ),
\end{align}
where $F_{\rm ch} = z - \hat{\tau} (2+\hat{a}) g^{\rm ch} $. In the case of \eqref{wish}, substituting
(\ref{kG}) in (\ref{chfin})  gives (\ref{ggW}).

With (\ref{ggW}) of interest in mathematical statistics and thus as a stand alone result, let us show
the method of characteristics discussed in Section \ref{S2} can be used to solve the partial differential equation
(\ref{wish}) directly.
We first note \eqref{wish} is an instance of \eqref{geng} with replacements $G = g^{\rm W}$ and $A=\frac{1}{2},B=\hat{a} + 2zg^{\rm W},C=-(g^{\rm W})^2$. The ODEs (\ref{cond}) and (\ref{chars}) read
\begin{align}
	\frac{\partial}{\partial \beta} \hat{\tau}(\alpha,\beta) & = \frac{1}{2}, \\
	\frac{\partial}{\partial \beta} z(\alpha,\beta) & = \hat{a} + 2 z(\alpha,\beta) g^{\rm W}(\alpha,\beta), \\
	\frac{\partial}{\partial \beta} g^{\rm W}(\alpha,\beta) & = - g^{\rm W}(\alpha,\beta)^2, 
\end{align}
where $(\alpha,\beta)$ are the transformed variables $(z,\hat{\tau})$ and the abbreviated notation $g^{\rm W}(\alpha,\beta) = g^{\rm W}(z(\alpha,\beta);\hat{\tau}(\alpha,\beta))$. The solution to these ODE's with initial conditions $z(\alpha,0) = \alpha$, $\hat{\tau}(\alpha,0) = 0$ and $g^{\rm W}(\alpha,0)=g^{\rm W}_{0}(\alpha)$ is
\begin{align}
	\hat{\tau}(\alpha,\beta) & = \frac{\beta}{2}, \label{wish1}\\
	z(\alpha,\beta) & = \hat{a} \beta(1+g^{\rm W}_{0}(\alpha) \beta) + \alpha(1+g^{\rm W}_{0}(\alpha) \beta)^2, \label{wish2}\\
	g^{\rm W}(\alpha,\beta) & = \frac{g^{\rm W}_{0}(\alpha)}{1+g^{\rm W}_{0}(\alpha)\beta}. \label{wish3}
\end{align}
Since to proceed requires purely algebraic operations only, we suppress all of the arguments in what follows. We find a formula for $g_{W,0}$ from the last equation,
\begin{align}
\label{gw0}
	g^{\rm W}_{0} = \frac{g^{\rm W}}{1-g^{\rm W} \beta},
\end{align}
and substitute it into \eqref{wish2} to obtain the formula for $\alpha$,
\begin{align}
\label{walpha}
	\alpha = z (1-g^{\rm W}\beta)^2 - \hat{a} \beta (1-g^{\rm W}\beta).
\end{align}
Next we turn to \eqref{gw0} and find 
\begin{align}
	g^{\rm W} = (1-g^{\rm W} \beta) g^{\rm W}_{0}
\end{align}
which, including the formula \eqref{walpha} for $\alpha$ and \eqref{wish1}, gives an implicit solution of \eqref{wish}
\begin{align}
\label{wishartsol0}
	& g^{\rm W}(z;\hat{\tau}) = \left (1-2\hat{\tau}g^{\rm W}(z;\hat{\tau}) \right )  g^{\rm W}_{0}\left( z \left (1-2\hat{\tau}g^{\rm W}(z;\hat{\tau}) \right )^2 - 2\hat{a}\hat{\tau} \left (1- 2\hat{\tau}g^{\rm W}(z;\hat{\tau}) \right ) \right ).
\end{align} 
Since
\begin{align}
	g^{\rm W}_{0}(z) = {\rm PV} \int_{I_+^{(0)}} \frac{\rho^{\rm W, (0)}(\mu)d\mu}{z-\mu}
\end{align}
with $I_+^{(0)} \subset R^+$ and $\rho^{\rm W, (0)}(\mu)$ is normalized to unity when integrating over $\mu>0$
we see that (\ref{ggW}) follows.

%%%%%%%%%%%%%%%%%%%%%%%%%%%%%%%%%%%%%%%%%%%%%%%%%%%%%%%%%%

\section{Circular and Jacobi ensembles}\label{S4}
\subsection{Circular ensembles}
We now turn our attention to Smoluchowski-Fokker-Planck type dynamics of circular ensembles. These models arise when considering random (symmetric if $\beta=1$, unrestricted if $\beta = 2$ or selfdual if $\beta=4$) unitary matrices $U$ of size $N\times N$ distributed according to the Haar measure; see e.g.~\cite[Ch.~2]{Fo10}.
 The Jacobian in this case reads
\begin{align}
	J = \prod_{1<i<j<N} |e^{i\phi_i} - e^{i\phi_j}|^\beta,
\end{align}
so that eigenvalues $e^{i\phi_i}$ lie on a unit circle. The diffusion is introduced based on the parametrization of $U$ in terms of exponent of an Hermitan matrix. The joint eigenvalue PDF satisfies  \eqref{FP} with a drift term 
\begin{align}
	W = - \sum_{1<i<j<N} \log |e^{i\phi_i} - e^{i\phi_j}|
\end{align}
and with the replacements $\{ \lambda_i \} \mapsto \{ \phi_i \}$.; see e.g.~\cite[\S 11.2.1]{Fo10}.

The log-gas system occupies the domain $\phi \in (-\pi,\pi]$ and is described by one and two body potentials
\begin{align}
	V_1(\phi) = 0, \quad V_2(\phi,\theta) = - \log |e^{i\phi} - e^{i\theta}|.
\end{align}
For the spectral density $\rho^{\circ}$
\begin{align}
	\rho^{\circ}(\phi;\tau) = \frac{1}{N} \left < \sum_{i=1}^N \delta (\phi - \phi_i) \right >_{P_\tau}
\end{align}
in the global regime where $\hat{\tau} = N \tau$, the hydrodynamical equation \eqref{H} is equal to
\begin{align}
\label{circ}
	\frac{\partial \rho^{\circ}(\phi;\hat{\tau})}{\partial \hat{\tau}} = - \frac{\partial}{\partial \phi}\left [ \rho^{\circ}(\phi;\hat{\tau}) \frac{\partial}{\partial \phi} \left ( \int_{-\pi}^\pi d \phi' \log |e^{i\phi} - e^{i\phi'}| \rho^{\circ}(\phi';\hat{\tau}) \right ) \right ] .
\end{align}
We define a circular Hilbert transform as
\begin{align}
\label{hilbc}
	\mathcal{H}_{\circ}[f](\phi) := \frac{1}{2} {\rm PV} \int_{I} d\phi' \cot \left(\frac{\phi-\phi'}{2} \right) f(\phi'), \qquad \phi \in \bar{I},
\end{align}
where $\bar{I} \subset (-\pi, \pi]$
and, since $\partial_x \log |e^{ix} - e^{iy}| = \frac{1}{2}\cot \left ( \frac{x-y}{2} \right )$, the equation \eqref{circ} is expressed as
\begin{align}
	\frac{\partial \rho^{\circ}(\phi;\hat{\tau})}{\partial \hat{\tau}} = - \frac{\partial}{\partial \phi}\left ( \rho^{\circ}(\phi;\hat{\tau}) \mathcal{H}_{\circ}[\rho^{\circ}(\cdot;\hat{\tau})](\phi) \right ) .
\end{align}
To arrive at the final evolution equation, we introduce a circular Green's function 
\begin{align}
	G^{\circ}(z;\hat{\tau}) = \frac{1}{2} \int_{\bar{I}} \cot \left ( \frac{z-y}{2} \right ) \rho^{\circ}(y;\hat{\tau})
	\, dy , 
\end{align}
which also satisfies \eqref{W} and \eqref{gpm} with replacements $G \mapsto G^{\circ}$ and $\rho \mapsto \rho^{\circ}$. Based on these properties, we again
find  the complex Burger's equation
\begin{align}
\label{circhydr}
	\frac{\partial}{\partial \hat{\tau}} G^{\circ}(z;\hat{\tau}) + G^{\circ}(z;\hat{\tau}) \frac{\partial}{\partial z} G^{\circ}(z;\hat{\tau}) = 0,
\end{align}
which is formally in the same form as the Gaussian case \eqref{G2}. We can thus apply the same techniques to conclude that the solution of the initial value problem for this equation ---  initial spectral density $\rho^{\circ,(0)}$  --- reads
\begin{align}\label{ss2}
	G^{\circ}(z;\hat{\tau}) = \int_{\bar{I}^{(0)}} \rho^{\circ,(0)}(\mu) \cot \left ( \frac{z - \hat{\tau} G^{\circ}(z;\hat{\tau}) - \mu}{2} \right ) \,  d\mu,
\end{align}
where $\bar{I}^{(0)}$ is the initial support of $\rho^{\circ,(0)}$.
The hydrodynamical equation (\ref{circ}) was first derived by Pandey and Shukla \cite[Eq.~(59)]{PS91},
using the hierarchy of equations satisfied by the dynamical correlation functions. The general solution
(\ref{ss2}) is given in  \cite[Eq.~(63)]{PS91}. Our main point here is therefore not a new result, but rather
a common theme, namely the macroscopic hydrodynamical equation (\ref{H}).

The particular case  $\rho^{\circ,(0)}(\mu) = \delta(\mu)$ was studied in the context of QCD by 
\cite{BLAIZOTNOWAK,BLAIZOTNOWAK1}. Even though there is no closed form solution of (\ref{ss2}), several analytic features
can be exhibited, 
including an effect analogous to that of the Gaussian ensemble evolution
with initial condition (\ref{Ga3}): at a critical value to $\hat{\tau}$ two spectrum edges collide here being the
left and right edges of the single interval of support.

%%%%%%%%%%%%%%%%%%%%%%%%%%%%%%%%%%%%%%%%%%%%%%%%%%%%%%%%%%

\subsection{Jacobi ensembles in trigonometric variables}
We now move to the example of Jacobi ensembles. Consider a unitary (symmetric for $\beta=1$, unconstrained by $\beta = 2$ or self dual for $\beta=4$) matrix $S$  of size $(n+m) \times (n+m)$ with $n\geq m$, divide it into 4 blocks 
\begin{align}
	S = \left ( \begin{matrix} r_{n\times n} & t'_{n\times m} \\t_{m\times n} & r'_{m\times m} \end{matrix} \right ),
\end{align}
and investigate singular values of sub-block $t'$. The corresponding Jacobian of this ensemble (see \cite[\S 11.2.3]{Fo10}) reads
\begin{align}
\label{jacjac}
	J = \prod_{j=1}^m (\lambda_j^2)^{\frac{\beta a'}{2}} \prod_{1\leq i<j\leq m} |\lambda_i^2 - \lambda_j^2|^\beta,
\end{align}
where $a' = n-m + 1 - \frac{1}{\beta}$ and $\lambda_i \in (0,1)$ denote non-zero singular values of $t'$. 

To obtain a Smoluchowski-Fokker-Planck equation for the joint PDF, we introduce new variables $\lambda_i = \sin \frac{\phi_i}{2}$, with $\phi_i\in (0,\pi)$. As was demonstrated in \cite[\S 11.2.3]{Fo10}, the new variables $\{ \phi_i \}$ permit an evolution \eqref{FP} with a drift term 
\begin{align}
\label{wj}
	W = - \frac{a'}{2} \sum_{i=1}^m \log \sin^2  \frac{\phi_i}{2} - \frac{b'}{2} \sum_{i=1}^m \log \cos^2  \frac{\phi_i}{2} - \sum_{1\leq j<k \leq m} \log \left |\sin^2 \frac{\phi_j}{2} - \sin^2 \frac{\phi_k}{2} \right |,
\end{align}
with $b' = \frac{1}{\beta}$ and replacements $\{ \lambda_i \} \to \{ \phi_i \}, N \to m$. When compared to the Jacobian \eqref{jacjac}, an extra $b'$ term arise by transforming the measure $d\lambda_i = \frac{1}{2} \cos \frac{\phi_i}{2} d\phi_i$.
We read off one- and two body interactions from \eqref{wj},
\begin{align*}
	V_1(\phi) & = - \frac{a'}{2} \log \sin^2  \frac{\phi}{2} - \frac{b'}{2} \log \cos^2  \frac{\phi}{2}, \qquad V_2(\phi,\phi') = - \log \left |\sin^2 \frac{\phi}{2} - \sin^2 \frac{\phi'}{2} \right |,
\end{align*}
and rewrite the latter
\begin{align}
	V_2(\phi,\phi') =- \ln \left |\sin \left (\frac{\phi - \phi'}{2} \right ) \right | - \ln \left | \sin \left (\frac{\phi + \phi'}{2} \right ) \right | ,
\end{align}
so that the $\phi \to - \phi$ symmetry is evident. Accordingly, we form a spectral density of the form
\begin{align}
	\rho^{\rm J}(\phi;\tau) = \frac{1}{m} \left < \sum_{i=1}^m \left (\delta (\phi - \phi_i) + \delta (\phi + \phi_i) \right ) \right >_{P_\tau},
\end{align}
normalized to unity when integrated over $\phi \in (0,\pi)$ and even in $\phi$. Both the interaction term and spectral density has features present in the chiral spectral density $\rho^{\rm c}$ \eqref{rhochdef} and the two body potential term \eqref{W9}. The current \eqref{H} driving the time evolution of $\rho^{\rm J}$ in the large $m$ limit reads
\begin{align}
	J_{\rm J}(\phi;\hat{\tau}) & = \rho^{\rm J}(\phi;\hat{\tau}) \frac{\partial}{\partial \phi} \left ( \frac{\hat{a}}{2} \log \sin^2 \frac{\phi}{2} + \int_{0}^\pi d\phi' \log \left |\sin^2 \frac{\phi}{2} - \sin^2 \frac{\phi'}{2} \right | \rho^{\rm J}(\phi';\hat{\tau}) \right ) \nonumber \\
	& = \rho^J(\phi;\hat{\tau}) \left ( \frac{\hat{a}}{2} \cot \frac{\phi}{2} + \frac{1}{2} \int_{-\pi}^\pi d\phi' \cot \frac{\phi - \phi'}{2} \rho^J(\phi';\hat{\tau}) \right ),
\end{align}
where the $\hat{a} = \frac{n}{m} - 1$, $b'$ term has dropped out as subleading in the large $m$ limit and 
the rescaled time parameter reads $\hat{\tau} = m \tau$. The hydrodynamic equation $	\frac{\partial}{\partial \hat{\tau}} \rho^{\rm J}(\phi;\hat{\tau}) = - \frac{\partial}{\partial \phi} J_{\rm J}(\phi;\hat{\tau})$ reads
\begin{align}
	\frac{\partial}{\partial \hat{\tau}} \rho^{\rm J}(\phi;\hat{\tau}) = - \frac{\partial}{\partial \phi} \Big ( \rho^{\rm J}(\phi;\hat{\tau}) \mathcal{H}_\circ[\hat{a} \delta(\cdot) + \rho^{\rm J}(\cdot;\hat{\tau})](\phi) \Big ),
\end{align}
where the Hilbert transform $\mathcal{H}_\circ$ was already defined in \eqref{hilbc}. By using the properties of Green's function 
\begin{align}\label{GJ}
	G^{\rm J}(z;\hat{\tau}) = \frac{1}{2} \int_{\bar{I}_+}\Big ( \cot \left ( \frac{z-y}{2} \right )
	+   \cot \left ( \frac{z+y}{2} 
	\right )\Big ) \rho^{\circ}(y;\hat{\tau}) 
	\, dy , 
\end{align}
with $\bar{I}_+ \subset (0,\pi]$, we repeat the derivation of the complex Burgers equation (\ref{circhydr}) and obtain 
\begin{align}
\label{jacg}
	\frac{\partial}{\partial \hat{\tau}} G^{\rm J}(z;\hat{\tau}) + \left ( \frac{\hat{a}}{2} \cot \frac{z}{2} + G^{\rm J}(z;\hat{\tau}) \right ) \frac{\partial}{\partial z} G^{\rm J}(z;\hat{\tau}) - \frac{\hat{a}}{4} \frac{G^{\rm J}(z;\hat{\tau})}{\sin^2 z/2} = 0.
\end{align}
The equation has the same structure as the chiral Gaussian equation \eqref{He}, and in
fact reduces to that equation for small $z$. The underlying log-gas setup has therefore the same features --- it consists of a fixed particle at $\phi=0$ of charge $\hat{a}$ and two mirror-like clouds for $\phi \in (-\pi,0)$ and $\phi \in (0,\pi)$ respectively. In the special case of vanishing charge $\hat{a} = 0$, the resulting equation \eqref{jacg} coincides exactly with \eqref{circhydr} obtained for the circular ensembles.   
%%%%%%%%%%%%%%%%%%%%%%%%%%%%%%%%%%%%%%%%%%%%%%%%%%%%%%%%%%

\section{Collective variables}\label{S5}
Collective variables is another approach to obtain the hydrodynamic equations \eqref{G2} and \eqref{He}. 
The idea of collective variables
was first  introduced in plasma physics \cite{BOHM} and extensively applied to gauge theories \cite{JEVICKISAKITA}.
Besides rederiving the aforementioned hydrodynamical equations, this method is suitable for obtaining asymptotic formulas for group integrals of Harish-Chandra/Itzykson--Zuber and Berezin--Karpelevich type. The former relate to the Gaussian ensembles whereas the latter appear in the chiral Gaussian ensembles.
This relationship is the reason why we focus only on these two cases in this section, and don't consider the
circular or Jacobi spectral evolutions.

\subsection{Collective variables method}
In the present context, one proceeds   by transforming the Smoluchowski-Fokker-Planck equation \eqref{FP} to new "collective" type variables $\hat{\lambda}$,
\begin{align}
\label{transf}
	\lambda_i \to \hat{\lambda}_j(\{\lambda \}), \qquad \{ \lambda \} = (\lambda_1 , ... ,\lambda_N)
\end{align}
where $i=1...N, j=1...N'$. These new degrees of freedom should a) use the symmetries of the system and b) have a well defined large $N$ limit. Typically $N' \to \infty$ from the beginning, and thus the particle system is
treated as a fluid, so that the change is not bijective at least before taking the large $N$ limit. For the
special case $N' = N$ and $N$ finite, the method corresponds  to a bona fide variable change and was  recently studied in the present context in
\cite{SMILANSKY}. The non-uniqueness of the collective variables means the aim is not
an exact description in all regimes. However, since the new degrees of freedom conserve the symmetries, one expects to correctly reproduce certain macroscopic properties.

Consider a general transformation $\hat{\lambda}(q;\{\lambda \})$ with $i$ index promoted to a variable $q$ (i.e. $N' \to \infty$) in a fluid approximation.
This continuous case introduces functional analysis by which the transformed
Smoluchowski-Fokker-Planck equation 
\begin{align}
\label{FP2}
\partial_\tau \pi_\tau(\{\lambda \}) = L(\{\lambda \})\pi_\tau(\{ \lambda \}) , \qquad L(\{ \lambda \}) = \frac{1}{\beta} \sum_{i=1}^N \frac{\partial^2 }{\partial {\lambda_i}^2}  - \sum_{i=1}^N \frac{\partial W}{\partial {\lambda_i}} \frac{\partial }{\partial {\lambda_i}}, 
\end{align}
obtained by writing  $p_\tau = \exp(-\beta W) \pi_\tau$ in \eqref{FP},
is transformed to a functional differential equation. 

%The rescaled joint PDF
%\begin{align}
%\pi_\tau = \int (U^\dagger dU) P_\tau(X^{(0)},U L U^\dagger)
%\end{align}
%can be defined in terms of $P_\tau$ introduced in \eqref{P1}. 

According to \eqref{transf}, the function of $\{ \lambda \}$ becomes a functional in the $q$ variables $\pi_\tau(\{ \lambda \}) = \hat{\pi}_\tau[\hat{\lambda}(q;\{\lambda \})]$. Moreover, the Laplace-Beltrami operator $L$ is re-expressed by the appropriate continuous chain rule
\begin{align}
	\frac{\partial}{\partial {\lambda_j}} = \sum_{i=1}^{N'} \frac{\partial{\hat{\lambda}_i}(\{\lambda \})}{\partial \lambda_j} \frac{\partial}{\partial \hat{\lambda}_i} \qquad \stackrel{"N'\to \infty"}{\longrightarrow} \qquad \frac{\partial}{\partial {\lambda_j}} = \int dq \frac{\partial{\hat{\lambda}}(q;\{\lambda \})}{\partial \lambda_j} \frac{\delta}{\delta \hat{\lambda}(q)},
\end{align}
where the $\hat{\lambda}(q)$ are the new variables just as $\hat{\lambda}_i$ in the discrete case. The 
transformed operator $\hat{L} = \hat{K} + \hat{V}$ reads
\begin{align}
	& \hat{K} = \frac{1}{\beta} \int dq \sum_{i=1}^N \frac{\partial^2 \hat{\lambda}(q)}{\partial \lambda_i^2}  \frac{\delta}{\delta \hat{\lambda}(q)} + \frac{1}{\beta} \int dp dq  \sum_{i=1}^N \frac{\partial \hat{\lambda}(q)}{\partial \lambda_i} \frac{\partial \hat{\lambda}(p)}{\partial \lambda_i} \frac{\delta^2}{\delta \hat{\lambda}(p) \delta \hat{\lambda}(q)}, \\
	& \hat{V} = - \int dq \left ( \sum_i \frac{\partial W}{\partial \lambda_i} \frac{\partial \hat{\lambda}(q)}{\partial \lambda_i} \right ) \frac{\delta}{\delta \hat{\lambda}(q)},
\end{align}
where we supressed the $\{ \lambda \}$ dependence in the coefficients. The transformed 
Smoluchowski-Fokker-Planck equation \eqref{FP2} is then
\begin{align}
 \partial_\tau \hat{\pi}_\tau[\hat{\lambda}] = \left ( \hat{K}[\hat{\lambda}] + \hat{V}[\hat{\lambda}] \right ) \hat{\pi}_\tau[\hat{\lambda}] .
\end{align}
%%%%%%%%%%%%%%%%%%%%%%%%%%%%%%%%%%%%%%%%%%%%%%%%%%%%%%%%%%
\subsection{Gaussian ensembles}
\label{s4.1}
 In the case of Gaussian ensembles, the drift term $W$ is given by \eqref{W} and the collective variable  
\begin{align}\label{cvG}
	\hat{\lambda}(q;\{\lambda \}) = \sum_{i=1}^N \delta (q - \lambda_i)
\end{align}
is the (non-averaged) one-point correlation function (see \eqref{rhodef}). This choice is consistent with condition a) mentioned in the introduction to this section --- it conserves the eigenvalue exchange symmetry. We calculate the kinetic part $\hat{K}$ with the help of the formula $\partial_{\lambda_i} \hat{\lambda} = - \partial_q \delta ( q - \lambda_i)$, and the potential part $\hat{V}$ using
\begin{align*}
& \frac{1}{\lambda_i - \lambda_j} = \text{PV} \int_I d\mu \, \frac{1}{\lambda_i-\mu} \delta(\lambda_j - \mu), \\
& \sum_{i\neq j} \delta (p - \lambda_i) \delta (q - \lambda_j) = \hat{\lambda}(p)\hat{\lambda}(q) - \delta (p - q) \hat{\lambda}(p).
\end{align*}
We set an ansatz for the leading large $N$ form of the joint PDF
\begin{align}
\label{ansatz1}
\hat{\pi}_\tau = \exp \left (-\frac{\beta}{2} N^2 S_\tau \right ), 	
\end{align}
we find that  the new functional $S_\tau$ satisfies the evolution equation
\begin{align}
\label{Seq}
	\partial_\tau S_\tau  = & \int dp \, \hat{\lambda}(p) \left ( \frac{1}{\beta} \frac{\partial^2}{\partial p^2} \left ( \frac{\delta S_\tau}{\delta \hat{\lambda}(p)} +\frac{\delta^2 S_\tau}{\delta \hat{\lambda}(p)^2} \right ) - \mathcal{H}[\delta](0) \frac{\partial}{\partial p} \frac{\delta S_\tau}{\delta \hat{\lambda}(p)}\right ) + \nonumber \\
	& - \int dp \, \hat{\lambda}(p) \left ( \frac{N^2}{2} \left ( \frac{\partial}{\partial p} \frac{\delta S_\tau}{\delta \hat{\lambda}(p)} \right )^2 - \mathcal{H}[\hat{\lambda}](p) \frac{\partial}{\partial p} \frac{\delta S_\tau}{\delta \hat{\lambda}(p)} \right ),
\end{align} 
where $\mathcal{H}[f]$ denotes the Hilbert transform \eqref{Ht} with supressed argument. 

We perform the large $N$ limit of \eqref{Seq} by rescaling both the time $N\tau= \hat{\tau}$ and the collective variable $\hat{\lambda} = N\rho$. In this limit, the first term on the RHS is subleading in $N$ in comparison to the second and the time derivative on LHS. Ignoring this term,
we  obtain an equation for $S_{\hat{\tau}}$ in the Hamilton-Jacobi form,
\begin{align}
	\partial_{\hat{\tau}} S_{\hat{\tau}} + \int dp \, \rho(p) \left [ \frac{1}{2} \left ( \frac{\partial}{\partial p} \frac{\delta S_{\hat{\tau}}}{\delta \rho(p)} \right )^2 - \mathcal{H}[\rho](p) \frac{\partial}{\partial p} \frac{\delta S_{\hat{\tau}}}{\delta \rho(p)} \right ] = 0,
\end{align}
where the position variable is $\rho(p)$ and the conjugate momentum reads $\Delta(p) = \frac{\delta S_{\hat{\tau}}}{\delta \rho(p)}$. This allows $S_{\hat{\tau}}$ to be interpreted as an action evaluated on a physical trajectory between $\rho(p;\hat{\tau}=0)$ and $\rho(p;\hat{\tau})$. The resulting Hamiltonian $H = \int dp \, \rho \left ( \frac{1}{2} (\partial_p \Delta )^2 - \mathcal{H}[\rho] \partial_p\Delta \right )$ needs a minor reformulation since it contains a problematic Hilbert transform term. To this end, we invoke a canonical change of variables $(\rho,\Delta) \to (\rho', \Delta' = \Delta + C)$ with $C$ dependent only on $\rho$. This change leaves the Hamiltonian unaltered i.e. $H'[\rho',\Delta'] = H[\rho', \Delta'-C[\rho']]$ and the action picks up a boundary term 
\begin{align}
\label{sb}
S'_{\hat{\tau}} = S_{\hat{\tau}} + T_{|\hat{\tau}} - T_{|0},
\end{align}
where the subscripts denote boundary terms evaluated at initial $\hat{\tau}=0$ and final time $\hat{\tau}$. The generating function $T$ is found to be
\begin{align}
	T = -\frac{1}{2} \int dp dq \,  \rho(p) \rho(q) \ln |p-q|, \quad \frac{\delta T}{\delta \rho(p)} = C.
\end{align}
The transformed Hamiltonian $H'$ is  
\begin{align}
\label{s0}
	H'[\rho',\Delta'] = \frac{1}{2} \int dp \, \rho'(p) \left [ \left ( \partial_p \Delta'(p) \right )^2 - \left ( \mathcal{H}[\rho'](p) \right )^2 \right ]
\end{align}
for which the second term is reexpressed in term of $\rho$ as
\begin{align}
\label{id1}
	\int dp \, \rho'(p) (\mathcal{H}[\rho'](p))^2 = \frac{\pi^2}{3} \int dp \, \rho'(p)^3 .
\end{align}
This identity is proved using the properties of the Hilbert transform 
\begin{align*}
&	\int f \mathcal{H}[g] = - \int g \mathcal{H}[f], \qquad 2 \mathcal{H}[f \mathcal{H}[f]] = (\mathcal{H}[f])^2 - \pi^2 f^2 ,
\end{align*}
valid for sufficiently well-behaved functions $f,g$ \cite{KING}. From now on we drop the prime indices and the Hamiltonian \eqref{s0} is finally
\begin{align}
H[\rho,\Delta] = \frac{1}{2} \int dp \, \rho \left ( (\partial_p \Delta)^2 - \frac{\pi^2}{3} \rho^2 \right ),
\end{align}
with corresponding action
\begin{align}\label{sA1}
S_{\hat{\tau}}  =  \frac{1}{2} \int_0^{\hat{\tau}} d \hat{t} \int dp \, \rho \left ( (\partial_p \Delta)^2 +\frac{\pi^2}{3} \rho^2 \right ),
\end{align}
chosen so that $S_{\hat{\tau}} |_{\hat{\tau}=0} = 0$ (otherwise $S_{\hat{\tau}}$ is unique only up to an additive constant).
By the Hamilton equations $\partial_{\hat{\tau}} \rho = \frac{\delta H}{\delta {\Delta}},
	 \partial_{\hat{\tau}} {\Delta} = -\frac{\delta H}{\delta \rho}$, the equations of motion read
\begin{align}
	& \partial_{\hat{\tau}} \Delta + \frac{1}{2} ( \partial_p \Delta )^2 = \frac{\pi^2 }{2} \rho^2, \nonumber \\
 & \partial_{\hat{\tau}} \rho + \partial_p (\rho \partial_p \Delta ) = 0.
\end{align}
Upon defining $G_\pm = \mp i \pi \rho + \partial_p\Delta$ (cf. \eqref{gm}), these formulas are exactly the complex Burger's equation \eqref{G2}. The construction goes similarly as before --- equations \eqref{gpm} for $G_\pm$ are defined on the real line and induce a complex structure due to analytic properties of $G$. 

This is a well-known result of Matytsin \cite{MATYTSIN}, reproduced also by other authors \cite{GUIONNET2, BOUCHAUD}. Here we show how additionally the joint PDF function $\hat{\pi}$ is asymptotically expressed in terms of an action related to the hydrodynamical system. 
%%%%%%%%%%%%%%%%%%%%%%%%%%%%%%%%%%%%%%%%%%%%%%%%%%%%%%%%%%
\subsection{Chiral Gaussian ensembles} 
\label{s4.2}
For the chiral case we make suitable replacements \eqref{R} and $W$ is defined in \eqref{W8}. The collective variable in this case is
\begin{align}
	\hat{x}(q,\{ x \}) = \sum_{i=1}^m \delta(q - x_i) + \delta (q + x_i) = \sum_{i=1}^m 2|q| \delta (q^2-x_i^2), 
\end{align}
a (non-averaged) one-point correlation function (see \eqref{rhochdef}). Our task is therefore to transform
(\ref{FP2}).
Because the derivation is parallel to the Gaussian case, we give only some partial results. 
%The rescaled joint PDF of is given by
%\begin{align}
%\label{P2resc}
%	\pi_\tau = \int (U^\dagger dU) (V^\dagger dV) P_\tau (Z^{(0)};ULV^\dagger)
%\end{align}
%where $P_\tau$ is defined in \eqref{P1a}.
In calculating the transformed Laplace operator $\hat{L}$, we use formulas
\begin{align*}
&	\frac{\partial}{\partial{x_j}} \left ( \sum_{i=1}^m 2|q| \delta (q^2 - x_i^2) \right ) = - \frac{|q|}{q} \frac{\partial}{\partial q} \Big (2|q| \delta (q^2 - x_j^2) \Big ), \\
&	\frac{1}{x_j} \delta(q^2 - x_j^2) = \text{PV} \int_{-\infty}^\infty \frac{d\mu \, \delta(\mu)}{|q| - \mu}  \delta(q^2 - x_j^2), \\
& \frac{2x_j}{x_j^2 - x_k^2} \delta (q^2 - x_j^2) = \text{PV} \int_{-\infty}^\infty \frac{d\mu 2|\mu| \, \delta(\mu^2 - x_k^2)}{|q| - \mu} \delta (q^2 - x_j^2).
\end{align*}
We make a large $m$ joint PDF ansatz
\begin{align}
	\hat{\pi}_\tau = \exp \left (-\frac{\beta}{4}m^2 S^{\rm c}_\tau \right )
\end{align}
which captures the rough degrees of freedom $\sim m^2$ and trivial $\beta$ dependence. The equation satisfied by $S^{\rm c}_\tau$ reads
\begin{align}
&	\partial_\tau S^{\rm c}_\tau = \int dp \, \hat{x}(p) \left ( \frac{1}{\beta} \frac{\partial^2}{\partial p^2} \left ( \frac{\delta S^{\rm c}_\tau}{\delta \hat{x}(p)} + 2 \frac{\delta^2 S^{\rm c}_\tau}{\delta \hat{x}(q)^2} \right ) + \mathcal{H}[\delta(.)](0) \frac{\partial}{\partial p} \frac{\delta S^{\rm c}_\tau}{\delta \hat{x}(p)} \right ) + \nonumber \\
&	- \int dp \, \hat{x}(p) \left ( \frac{m^2}{2} \left ( \frac{\partial}{\partial p} \frac{\delta S^{\rm c}_\tau}{\delta \hat{x}(p)} \right )^2 - \mathcal{H}[a'\delta(.) + \hat{x}(.)](p) \frac{\partial}{\partial p} \frac{\delta S^{\rm c}_\tau}{\delta \hat{x}(p)} \right ).
\end{align}
Now we perform a $m,n \to \infty$ limit with $n/m$ fixed. We set $\hat{x} = m \rho^{\rm c}, \hat{\tau} = m \tau$ and find the first term on RHS subleading wrt. the second and the time derivative. The equation for $S^{\rm c}_{\hat{\tau}}$ is again in the Hamilton-Jacobi form
\begin{align}
	\partial_{\hat{\tau}} S^{\rm c}_{\hat{\tau}} + \int dp \, \rho^{\rm c}(p) \left ( \frac{1}{2} \left ( \frac{\partial}{\partial p} \frac{\delta 
	S^{\rm c}_{\hat{\tau}}}{\delta \rho^{\rm c}(p)} \right )^2 - \frac{a'}{p} \frac{\partial}{\partial p} \frac{\delta S^{\rm c}_{\hat{\tau}}}{\delta \rho^{\rm c}(p)} - \mathcal{H}[\rho^{\rm c}](p) \frac{\partial}{\partial p} \frac{\delta S^{\rm c}_{\hat{\tau}}}{\delta \rho^{\rm c}(p)} \right ) = 0. 
\end{align}
With the conjugate momentum $\Delta^{\rm c} = \frac{\delta S^{\rm c}_{\hat{\tau}}}{\delta \rho^{\rm c}}$, we transform $(\rho^{\rm c},\Delta^{\rm c}) \to ((\rho^{\rm c})'=\rho^{\rm c},(\Delta^{\rm c})'=\Delta^{\rm c} + C^{\rm c})$ and the action is transformed accordingly
\begin{align}
(S^{\rm c}_{\hat{\tau}})'  = S^{\rm c}_{\hat{\tau}} + T^{\rm c}_{|\hat{\tau}} - T^{\rm c}_{|0}	
\end{align}
with a generating function
\begin{align}
	T^{\rm c} = - \frac{1}{2} \int_{-\infty}^{\infty} dp dq \, \log|p-q| \rho^{\rm c}(p) \rho^{\rm c}(q), \qquad \frac{\delta T^{\rm c}}{\delta \rho^{\rm c}} = C^{\rm c} .
\end{align}
Under this transformation, the new Hamiltonian reads
\begin{multline}
 H[\rho^{\rm c},\Delta^{\rm c}] = \frac{1}{2} \int dp \, \rho^{\rm c}(p) \bigg ( \Big ( \frac{\partial}{\partial p} \frac{\delta S^{\rm c}_{\hat{\tau}}}{\delta \rho^{\rm c}(p)} \Big )^2 \\ - \Big ( \mathcal{H}[\rho^{\rm c}](p)  \Big )^2 + \frac{2\hat{a}}{p} \Big ( \frac{\partial}{\partial p} \frac{\delta S^{\rm c}_{\hat{\tau}}}{\delta \rho^{\rm c}(p)} - \mathcal{H}[\rho^{\rm c}](p) \Big ) \bigg ),
\end{multline}
where $\hat{a} = \frac{n}{m} - 1$ and we dropped the primed indices. As before, the term quadratic in Hilbert transforms is given in \eqref{id1} whereas the linear term vanishes by two additional properties of $\mathcal{H}$ \cite{KING},
\begin{align}
&	\mathcal{H}\left [\frac{f(.)}{(.)} \right ](x) = \frac{\mathcal{H}[f]}{x} - \frac{\mathcal{H}[f](0)}{x}, \\
&	\mathcal{H}[f](0)=0,\quad \text{if} \quad f(x) = f(-x).
\end{align}  
Thus the final form of the Hamiltonian reads
 \begin{align}
	H[\rho^{\rm c},\Delta^{\rm c}] = \frac{1}{2} \int dp \, \rho^{\rm c} \left ( (\partial_p \Delta^{\rm c})^2 - \frac{\pi^2}{3} (\rho^{\rm c})^2 + \frac{2\hat{a}}{p} \partial_p \Delta^{\rm c} \right ),
\end{align}
and the action is therefore
\begin{align}\label{Sc2}
S^{\rm c}_{\hat{\tau}}  = 
\frac{1}{2}\int_0^{\hat{\tau}} d\hat{t} \int dp \, \rho^{\rm c} \left ( (\partial_p \Delta^{\rm c})^2 + \frac{\pi^2}{3} (\rho^{\rm c})^2 + \frac{2\hat{a}}{p} \partial_p \Delta^{\rm c} \right ).
\end{align}
The Hamilton equations of motion read
\begin{align}
	& \partial_{\hat{\tau}} \Delta^{\rm c} + \frac{1}{2} ( \partial_p \Delta^{\rm c} )^2 + \frac{\hat{a}}{p} \partial_p \Delta^{\rm c}  = \frac{\pi^2 }{2} (\rho^{\rm c})^2, \nonumber \\
 & \partial_{\hat{\tau}} \rho^{\rm c} + \partial_p (\rho^{\rm c} \partial_p \Delta^{\rm c} ) + \partial_p \left (\frac{\hat{a}}{p} \rho^{\rm c} \right )= 0.
\end{align}
We observe again how the chiral case reduces to Gaussian \eqref{G2} when $\hat{a} = 0$, and by defining $G^{\rm c}_\pm = \mp i \pi \rho^{\rm c} + \partial_p \Delta^{\rm c}$ we reclaim  \eqref{He} by the arguments elucidated previously. 
%%%%%%%%%%%%%%%%%%%%%%%%%%%%%%%%%%%%%%%%%%%%%%%%%%%%%%%%%%
\section{Asymptotic expansion of Harish-Chandra/Itzykson--Zuber and Berezin--Karpelevich integrals}\label{S6}
Collective variables were used by Matytsin \cite{MATYTSIN} to obtain large $N$ expansion of the celebrated Harish-Chandra/ Itzykson--Zuber integral formula. Later works looked at the same task from both mathematical \cite{GUIONNET2} and physical point of view \cite{BOUCHAUD}. In this section we comment on this standard result and afterwards use analogous working to compute an expansion for the Berezin--Karpelevich type integrals \cite{BK, GUHR} arising in the chiral Gaussian ensembles.
%%%%%%%%%%%%%%%%%%%%%%%%%%%%%%%%%%%%%%%%%%%%%%%%%%%%%%%%%%
\subsection{HCIZ-type integrals}
We consider an integral 
\begin{align}
\label{hciz}
	I_\beta(A,B) = \int (U^\dagger dU) \exp \left ( \frac{\beta N}{2} \text{Tr} (U A U^\dagger B ) \right ),
\end{align}
where matrices $A,B$ are diagonal of size $N\times N$ and $U$ are real orthogonal ($\beta=1$) or complex unitary ($\beta=2$). In the RMT context these integrals arise in connection with the Gaussian ensembles. For $\beta=2$ an exact formula exists, found independently by Charish--Handra \cite{CH} and Itzykson--Zuber \cite{IZ}. 

To obtain large $N$ asymptotic behaviour of \eqref{hciz}, we recall the definition \eqref{P1} of joint PDF 
\begin{align}
\label{pi}
	\pi_{\hat{\tau}} = \int (U^\dagger dU) P_{\hat{\tau}}(X^{(0)};ULU^\dagger) = \frac{1}{C_{N,\hat{\tau}}} e^{-\frac{\beta N}{4\hat{\tau}}\text{Tr} L^2 -\frac{\beta N}{4\hat{\tau}}\text{Tr} (X^{(0)})^2} I_\beta \left (\frac{L}{\sqrt{\hat{\tau}}},\frac{X^{(0)}}{\sqrt{\hat{\tau}}} \right ).
\end{align}
The traces in this expression can be given in term of the collective variable \eqref{cvG} as $\text{Tr} L^2 = N \int dp \, p^2 \rho(p;\hat{\tau})$ and $\text{Tr} (X^{(0)})^2 = N \int dp \, p^2 \rho(p;\hat{\tau}=0)$ in accordance to their role as an initial and final densities respectively. On the other hand, the asymptotic form of $\pi_\tau$ was found in Section \ref{s4.1} as
\begin{align}
	& \pi_{\hat{\tau}} \sim \exp \left ( -\frac{\beta}{2} N^2 (S_{\hat{\tau}} - T_{|\hat{\tau}} - T_{|0}) \right )
\end{align} 
using \eqref{ansatz1} and \eqref{sb}, and where we also added an arbitrary constant to the action $S_{\hat{\tau}} \to S_{\hat{\tau}} -2T_{|0}$ 
(recall the comment below (\ref{sA1})). The form of this constant is chosen so that $ \pi_{\hat{\tau}}|_{\hat{\tau} = 0}
 \sim \exp \left ( \beta N^2  T_{|0} \right )$.

To arrive at an asymptotic expression for \eqref{hciz}, we fix the time $\hat{\tau} = 1$ and rename the final $\rho(p,\hat{\tau}=1) = \rho_f(p)$ and initial $ \rho(p,\hat{\tau}=0) = \rho_i(p)$ densities 
\begin{align}
\label{hcizas}
	I_\beta(\sigma,\alpha) & \sim C_{N,\hat{\tau}=1} \exp \left ( \frac{\beta}{2} N^2 \left [ - S_{\hat{\tau}=1} + \frac{1}{2} \int dp ~ p^2 (\rho_i(p) + \rho_f(p)) + \right . \right . \nonumber \\
	& \left . \left . + \frac{1}{2} \int dp dq \Big (  \rho_i(p) \rho_i(q) + \rho_f(p) \rho_f(q) \Big ) \ln|q-p| \right ] \right ).
\end{align}
Now the main difficulty lies in finding a physical path joining initial $\rho_i(p)$ and final $\rho_f(p)$ spectral densities and calculating the corresponding action $S_{\hat{\tau}}$, which is specified by (\ref{sA1}). The former problem has been solved from our workings in Sections 1 and 2 for the
initial condiitions (\ref{1.5a}) and (\ref{Ga3}), while the evaluation of $S_{\hat{\tau}}$ in the first of these is given in \cite{BOUCHAUD}.
For a discussion of analyticity properties of (\ref{hcizas}) see \cite{GGN11}.
%%%%%%%%%%%%%%%%%%%%%%%%%%%%%%%%%%%%%%%%%%%%%%%%%%%%%%%%%%
\subsection{Berezin-Karpelevich type integrals}
We now turn to the asymptotic formula for an integral of Berezin-Karpelevich type defined as
\begin{align}
J_{\beta} (A,B) = \int (U^\dagger dU) (V^\dagger dV)\exp \left ( \frac{\beta m}{4} \text{Tr} \left ( V A^\dagger U^\dagger B + B^\dagger U A V^\dagger \right )\right ),
\end{align}
where $A,B$ are $n\times m$ diagonal matrices and $U,V$ are real orthogonal ($\beta=1$) or complex unitary ($\beta=2$) matrices of sizes $n\times n$ and $m\times m$ respectively with $n \geq m$. These integrals arise in studying chiral/Wishart/Laguerre type ensembles. 

In the $\beta=2$ case, an exact formula was rediscovered in \cite{GUHR} and originally calculated by Berezin and Karpelevich \cite{BERKARP}. To obtain an asymptotic expression for $\beta=1,2$ we recall the chiral joint PDF \eqref{P1a},
\begin{align}
\label{rhs}
	\pi_{\hat{\tau}} & = \int (U^\dagger dU) (V^\dagger dV) P_{\hat{\tau}} (Z^{(0)};ULV^\dagger) 
	\nonumber \\ & = \frac{1}{C^{\rm c}_{N,\hat{\tau}}} e^{-\frac{\beta m}{4\hat{\tau}}\text{Tr} \left ( L^\dagger L + (Z^{(0)})^\dagger Z^{(0)} \right )} J_\beta \left (\frac{L}{\sqrt{\hat{\tau}}}, \frac{Z^{(0)}}{\sqrt{\hat{\tau}}} \right ) .
\end{align}
We introduce the normalized collective variables to the Gaussian terms $\text{Tr} L^\dagger L = \frac{m}{2} \int dq~q^2 \rho^{\rm c}(q;\hat{\tau}) $ and $\text{Tr} (Z^{(0)})^\dagger Z^{(0)}  = \frac{m}{2} \int dq ~ q^2 \rho^{\rm c}(q;\hat{\tau}=0)$. 
An asymptotic form of LHS was found in Section \ref{s4.2} as
\begin{align}
\label{lhs}
	\pi_{\hat{\tau}} \sim \exp \left ( -\frac{\beta m^2}{4} \left (S^{\rm c}_{\hat{\tau}}- T^{\rm c}_{|\hat{\tau}} - T^{\rm c}_{|0} \right ) \right )
\end{align}
along with adding a constant $S^{\rm c}_{\hat{\tau}} \mapsto S^{\rm c}_{\hat{\tau}} -2T^{\rm c}_{|0}$. By comparing \eqref{rhs} and \eqref{lhs}, for fixed time $\hat{\tau}=1$ we have an asymptotic formula
\begin{align}
	J_\beta\left (\rho_f^{\rm c},\rho_f^{\rm c} \right ) & \sim C^{\rm c}_{N,\hat{\tau}=1} \exp \left ( \frac{\beta}{4} m^2 \left [ - S^{\rm c}_{\hat{\tau=1}} + \frac{1}{2} \int dp ~ p^2 \left (\rho_i^{\rm c}(p) + \rho_f^{\rm c}(p) \right ) + \right . \right . \nonumber \\
	& \left . \left .  + \frac{1}{2} \int dp dq \left (  \rho_i^{\rm c}(p) \rho_i^{\rm c}(q) + \rho_f^{\rm c}(p) \rho_f^{\rm c}(q) \right ) \ln|q-p| \right ] \right ).
\end{align}
where initial and final densities are denoted as $\rho^{\rm c}(p;\hat{\tau}=0) = \rho^{\rm c}_i(p)$ and $\rho^{\rm c}(p;\hat{\tau}=1) = \rho^{\rm c}_f(p)$ respectively. As in the case of (\ref{hcizas}), we comment that
to obtain the asymptotic formula for prescribed initial $\rho^{\rm c}_{i}$ and final $\rho^{\rm c}_{f}$ densities, it is necessary
 to evaluate the action $S^{\rm c}_{\tau}$ on a physical trajectory connecting these two spectral densities.
 We note that such trajectories are given for particular initial conditions below (\ref{chfin}).
 
 \subsection*{Acknowledgements}
 The work of PJF was supported by the Australian Research Council discovery project grant
 DP140102613 and by the ARC Centre of Excellence for Mathematical and Statistical Frontiers. JG thanks Melbourne University for the warm hospitality during the preparation of this work and acknowledges the support of both the Grant DEC-2011/02/A/ST1/00119 of the National Centre of Science and the Australian Government Endeavour Fellowship.

\end{document}